\documentclass{aa}
\usepackage{txfonts}
\usepackage[comma,authoryear]{natbib}
\usepackage[latin1]{inputenc}

\usepackage[dvips]{graphicx}

\def\bbbc{{\mathchoice {\setbox0=\hbox{$\displaystyle\rm C$}\hbox{\hbox
to0pt{\kern0.4\wd0\vrule height0.9\ht0\hss}\box0}}
{\setbox0=\hbox{$\textstyle\rm C$}\hbox{\hbox
to0pt{\kern0.4\wd0\vrule height0.9\ht0\hss}\box0}}
{\setbox0=\hbox{$\scriptstyle\rm C$}\hbox{\hbox
to0pt{\kern0.4\wd0\vrule height0.9\ht0\hss}\box0}}
{\setbox0=\hbox{$\scriptscriptstyle\rm C$}\hbox{\hbox
to0pt{\kern0.4\wd0\vrule height0.9\ht0\hss}\box0}}}}
\def\bbbq{{\mathchoice {\setbox0=\hbox{$\displaystyle\rm
Q$}\hbox{\raise
0.15\ht0\hbox to0pt{\kern0.4\wd0\vrule height0.8\ht0\hss}\box0}}
{\setbox0=\hbox{$\textstyle\rm Q$}\hbox{\raise
0.15\ht0\hbox to0pt{\kern0.4\wd0\vrule height0.8\ht0\hss}\box0}}
{\setbox0=\hbox{$\scriptstyle\rm Q$}\hbox{\raise
0.15\ht0\hbox to0pt{\kern0.4\wd0\vrule height0.7\ht0\hss}\box0}}
{\setbox0=\hbox{$\scriptscriptstyle\rm Q$}\hbox{\raise
0.15\ht0\hbox to0pt{\kern0.4\wd0\vrule height0.7\ht0\hss}\box0}}}}
\def\bbbt{{\mathchoice {\setbox0=\hbox{$\displaystyle\rm
T$}\hbox{\hbox to0pt{\kern0.3\wd0\vrule height0.9\ht0\hss}\box0}}
{\setbox0=\hbox{$\textstyle\rm T$}\hbox{\hbox
to0pt{\kern0.3\wd0\vrule height0.9\ht0\hss}\box0}}
{\setbox0=\hbox{$\scriptstyle\rm T$}\hbox{\hbox
to0pt{\kern0.3\wd0\vrule height0.9\ht0\hss}\box0}}
{\setbox0=\hbox{$\scriptscriptstyle\rm T$}\hbox{\hbox
to0pt{\kern0.3\wd0\vrule height0.9\ht0\hss}\box0}}}}
\def\bbbz{{\mathchoice {\hbox{$\sf\textstyle Z\kern-0.4em Z$}}
{\hbox{$\sf\textstyle Z\kern-0.4em Z$}}
{\hbox{$\sf\scriptstyle Z\kern-0.3em Z$}}
{\hbox{$\sf\scriptscriptstyle Z\kern-0.2em Z$}}}}

\newcommand{\ie}{{\it i.e.}\ }

\newcommand{\BVF}{Brunt-V\"ais\"al\"a frequency}

\newcommand{\beq}{\begin{equation}}
\newcommand{\beqa}{\begin{eqnarray*}}
\newcommand{\beqan}{\begin{eqnarray}}
\newcommand{\greq}{\begin{equation}\left\{ \begin{array}{l}}
\newcommand{\egreq}{\end{array}\right. \end{equation}}
\newcommand{\nngreq}{\[\left\{ \begin{array}{l}}
\newcommand{\nnegreq}{\end{array}\right. \]}
\newcommand{\egreqn}[1]{\end{array}\right. \label{#1}\end{equation}}
\newcommand{\eeq}{\end{equation}} 
\newcommand{\eeqn}[1]{\label{#1}\end{equation}} 
\newcommand{\eeqa}{\end{eqnarray*}}
\newcommand{\eeqan}[1]{\label{#1}\end{eqnarray}}

\newcommand{\lp}{ \left(}
\newcommand{\rp}{ \right)}

\newcommand{\khi}{\chi}

\newcommand{\eps}{\varepsilon}

\renewcommand{\na}{ \vec{\nabla} }

\newcommand{\intsur}{ \int_{(S)}\! }
\newcommand{\intvol}{ \int_{(V)}\! }
\newcommand{\cth}{ \cos\theta }

\newcommand{\sth}{ \sin\theta }

\newcommand{\vu}{\vec{u}}

\newcommand{\es}{\vec{e}_s}

\newcommand{\ez}{\vec{e}_z}

\newcommand{\vn}{\vec{n}}

\newcommand{\vr}{\vec{r}}

\newcommand{\vy}{\vec{y}}

\newcommand{\vO}{\vec{\Omega}}

\newcommand{\dV}{d^3\vec{r}}

\newcommand{\demi}{\frac{1}{2}}

\newcommand{\dr}[1]{\frac{\partial  #1}{\partial r}}

\newcommand{\dnr}[1]{\frac{d  #1}{dr}}

\newcommand{\dz}[1]{\frac{\partial  #1}{\partial z}}

\renewcommand{\div}{\vec{\nabla}\cdot}

\newcommand{\eq}[1]{(\ref{#1})}

\begin{document}
\author{F. Espinosa Lara \and M. Rieutord}
\institute{
Laboratoire d'Astrophysique de Toulouse et Tarbes, UMR 5572, CNRS et
Université Paul Sabatier Toulouse 3, 14 avenue E. Belin, 31400 Toulouse,
France}

\title{The dynamics of a fully radiative rapidly rotating star enclosed within a spherical
box}
\date{\today}
\abstract{
Recent results from interferometry and asteroseismology require
models of rapidly rotating stars that are more and more precise.
}{
We describe the basic structure and the hydrodynamics of a fully
radiative  star as a preliminary step towards more realistic models of
rotating stars.
}{
We consider a solar mass of perfect gas enclosed in a spherical
container. The gas is self-gravitating and rotating, and is the seat of
nuclear heating, and heat diffusion is due to radiative diffusion with
Kramers type opacities. Equations are solved numerically with spectral
methods in two dimensions with a radial Gauss-Lobatto grid and
spherical harmonics.
}{
We computed the centrifugally flattened structure of such a star: the
von Zeipel model, which says that the energy flux is proportional to
the local effective gravity is tested. We show that it overestimates
the ratio of the polar to the equatorial energy flux by almost a factor 2. We
also determine the \BVF\ distribution and show that
outer equatorial regions in a radiative zone are convectively unstable
when the rotation is fast enough. We compute the differential rotation and
meridional circulation stemming from the baroclinicity of the star and
show that, in such  radiative zones, equatorial regions rotate faster
than polar ones.  The surface differential rotation is also shown to
reach a universal profile when rotation is slow enough (less than 36\%
of the breakup one), as long as viscosity and Prandlt number remain small.
}{
}

\maketitle

\section{Introduction}

Recent observations of nearby stars with optical or infrared
interferometers have opened a new window on rapidly rotating stars. The
work of
\cite{DKJVONA05,Zorecetal05,McAlisteretal05,petersonetal06a,petersonetal06b,KDS06}
and \cite{Aufdenbergetal06}
show that observational techniques are now able to constrain the surface
distribution of the energy flux of these stars along with their precise
shape on the background sky. Data are usually fit by a simple brightness
law derived from a Roche-von Zeipel model \cite[e.g.][]{DVJJA02} where it
is assumed that the shape of the star is determined by a Roche model (all
the mass is concentrated in the center), while the brightness distribution
obeys the von Zeipel law, namely the energy flux is proportional to
local effective gravity.  If such a model is efficient at giving a first
interpretation of the observation and especially getting an idea of the
orientation of the stellar rotation axis, as well as its angular velocity,
one now wishes to go beyond this simple model. This step is 
difficult, however, as it means building a self-consistent two-dimensional
model of a star and then being able to make it evolve till the assumed
age of the observed object.

Many attempts have been made to reach this goal, but most of them
have not gone beyond the construction of barotropic stars without any
dynamics or evolution \cite[see the recent work of][]{Roxburgh04,JMS05}.
The inclusion of fluid dynamics is, however, important for the models
to be physically self-consistent. Rotating stars are indeed baroclinic,
and fluid flows are present everywhere, even in radiative zones. Their
long-term effect is the mixing of elements and  the transport of angular
momentum \cite[see][]{zahn92}, so they cannot be ignored in the evolution of
rotating stars.

Previous work on models of baroclinic stars is rare. Only \cite{UE94,UE95}
constructed global 2D models of rapidly rotating stars with baroclinicity,
but their neglect of viscosity removed any meridional circulation and
left differential rotation undetermined. More recently, \cite{R06} used
a Boussinesq model (\ie a nearly incompressible fluid) for investigating
the baroclinic flows on a global scale. In this case, it could be shown
how viscosity determines the differential rotation and that this
rotation strongly depends on the Prandtl number.

The present work aims at describing a more realistic, physically
self-consistent,  approach of a rotating, fully radiative star. Of course this
is a model and real physics is still very simplified, but no arbitrary
assumption needs to be made. In a few words, our model can be described
as follows: we consider a perfect gas contained in a rigid sphere; the
gas is rotating and self-gravitating; the mass inside the sphere is
high enough so that the central temperature and density permit nuclear
reactions.  The gas is a viscous fluid conducting heat through radiative
diffusivity derived from Kramers type opacities. The bounding sphere allows
us to impose simple (but physically relevant) boundary conditions
on the gravitational potential, temperature, pressure, and velocity
field. Roughly speaking, these conditions specify (i) that the gravitational
potential connects to the one pervading vacuum, (ii) that temperature
connects to the temperature field of a medium with a constant absorption
coefficient radiating like a black body, (iii) that some radial stress can
be accepted by the bounding sphere, but (iv) that matter slips freely and
tangentially on the sphere. This set-up, which is similar to a laboratory
experiment, simplifies the numerical resolution of the equations by
allowing the use of spherical coordinates while the distribution of mass
is not spherically symmetric. Fortunately, the bulk properties of this
``star" do not depend very much on the physics imposed at the outer
boundary. The next step will be to replace the sphere by a spheroid
that fits an isobar perfectly; however, this step, which needs to use
spheroidal coordinates with multi-domains, is numerically challenging
and can be undertaken only on a physically understood system.

We have voluntarily removed the convection zones that need a generalization
of the mixing-length theory in two dimensions including the effects of
rotation. Here we focus on a rotating radiative star and examine the dynamical
properties of its baroclinic state.

The paper is organized as follows. We next describe our model
(Sect.~2) and the numerical methods we use for solving the equations
(Sect.~3). We then describe some tests of the model showing internal
accuracy and how they compare to one-dimensional models (Sect.~4). In
Sect.~5 results are presented and discussed, and conclusions follow.

\section{The model}

We consider a viscous, self-gravitating fluid rotating with a mean
angular velocity $\vec{\Omega}=\Omega\vec{e}_z$. The fluid is supposed
to be inside an immutable sphere of radius $R$, so the outer boundary
can be taken at $r=R$. This simplifies the problem by allowing
spherical coordinates to represent the variables, although the external
boundary does not fit an isobar.

In a steady configuration, the fluid is governed by the
equations

\begin {equation}
\label{dim_eq}
\left\{\begin{array}{l}
\displaystyle \Delta\phi=4\pi G\rho\\
\\
\displaystyle \rho T\vec{v}\cdot\nabla s=-\div\vec{F}+\varepsilon\\
\\
\displaystyle \rho\left(2\vec{\Omega}\times\vec{v}+\vec{v}\cdot\nabla\vec{v}\right)=
-\nabla P-\rho\nabla\left(\phi-\frac{1}{2}\Omega^2r^2\sin^2\theta\right)
+\vec{F}_v\\
\\
\displaystyle \div(\rho\vec{v})=0
\end{array}
\right.
\end{equation}
where $\phi$ is the gravitational potential, $\rho$ the density, $s$
the specific entropy, $T$ the temperature, $\vec{v}$ the fluid velocity
with respect to a frame rotating at angular velocity $\vO$. Also, $P$
is the pressure and $\varepsilon$ the nuclear energy generation rate per
unit volume.  For a compressible fluid with constant dynamical viscosity
$\mu$, the viscous force can be written as

\begin{equation}
\vec{F}_v=\mu\left(\Delta\vec{v}+\frac{1}{3}\nabla(\div\vec{v})\right).
\end{equation}
For the energy flux, we will consider only radiative transport,
\begin{equation}
\vec{F}=-\chi\nabla T
\end{equation}
where $\chi(\rho,T)$ is the radiative diffusivity. The equation of state
is that of an ideal gas mixed with photons; hence, we use

\begin{equation}
p=\mathcal{R}_M\rho T + \frac{a}{3}T^4
\end{equation}
where $\mathcal{R}_M$ is the gas constant divided by the mean molecular
weight and $a$ the radiation density constant.

\subsection{Stellar microphysics}

We assume that the fluid is in the conditions of a stellar plasma typical
of the sun's radiative zone. The nuclear energy generation rate per unit
volume has the form

\begin{equation}
\varepsilon=\varepsilon_0 X^2\rho^2T^{-2/3} \mathrm{e}^{-bT^{-1/3}}
\end{equation}
where $X$ is the hydrogen mass fraction. For the constants
$\varepsilon_0$ and $b$ we adopt the values of the CESAM code
\cite[][]{Morel97}, namely
\begin{equation}
b=3600\quad\mbox{and}\quad \varepsilon_0=8.37\,10^{10}\quad\mbox{(cgs)}
\end{equation}
for the pp-chain.

In order to describe the radiative transport of energy, we use the
opacity given by power laws:

\begin{equation}
\kappa=\kappa_0 T^{-\beta} \rho^\eta, \label{opa}
\end{equation}
so that the radiative diffusivity can be written as
\begin{equation}
\chi=\frac{16\sigma T^3}{3\kappa\rho}=\chi_0T^{\beta+3}\rho^{-\eta-1}.
\end{equation}
In particular we use $\beta=1.97541$, $\eta=0.138316$, and
\mbox{$\kappa_0=7.1548412\,10^{13}$ cgs}, following
some fitting formulae\footnote{The proposed formula is
$\frac{1}{\kappa}=\frac{1}{\kappa_i}+\frac{1}{\kappa_e}$
with $\kappa_i$ as in~\eq{opa} and
$\kappa_e=1.6236784\,10^{-33}\rho^{0.407895}T^{9.28289}$; we use only
the ``interior" opacity, leaving the ``envelope" one.} proposed by J.
Christensen-Dalsgaard for simple solar models \cite[e.g.][]{CDGR95}.

\subsection{Boundary conditions}
In order to solve the full system of equations, we must complete it with
boundary conditions. At the center of the star ($r=0$) we just need
to impose regularity on the solutions. At the outer boundary, \ie on
the rigid bounding sphere, we impose that the fluid slips freely
and thus use stress-free boundary conditions. Hence, we have

\beq
u_r=\sigma_{r\theta}=\sigma_{r\varphi}=0\qquad \mathrm{at}\quad r=R
\eeq
where $\sigma_{r\theta}$ and $\sigma_{r\varphi}$ are the horizontal
components of the stress on the outer sphere.

Since the sphere is rigid, it can support some normal stress, and pressure
is not constant on it any more than the density and temperature. As the
true surface of the star is outside the box, pressure needs to be fixed
somewhere. We thus impose the polar pressure. This condition
imposes the place where our container ``cuts" the star envelope.

The usual boundary conditions on the temperature are that the boundary
radiates like a black body. We thus write

\[ -\khi\dr{T} = \sigma T^4 \qquad \mathrm{at}\quad r=R.\]
Using the expression of radiative diffusivity, this condition is
equivalent to

\[ \dr{T} + \sigma_T T = 0\]
where $\sigma_T=3\rho\kappa/16$. Obviously, $\sigma_T$ depends on
latitude, but to avoid unnecessary complications (nonlinearities)
we take it as a constant. Physically, this means that at the boundary
the fluid meets a medium with constant absorption coefficient
radiating like a black body.

Finally, the boundary conditions on the gravitational potential are that
it matches the vacuum solution, which vanishes at infinity.

\subsection{Dimensionless equations}

We scale the physical quantities, so that the solutions are of order
unity. For that, we choose the free-fall time scale \mbox{$\lambda=(4\pi
G\rho_c)^{-1/2}$}, the radius of the container $R$ as the length scale, the
central values of temperature $T_c$ and density $\rho_c$ for the
corresponding scales, and the gas constant $\mathcal{R}_M$ for the entropy
scale.  With dimensionless variables, the equations are now written as

\begin {equation}
\label{nondim_eq}
\left\{\begin{array}{l}
\displaystyle \Delta\phi=\rho\\
\displaystyle \frac{\mathcal{P}}{E}\rho T\vec{u}\cdot\nabla s=\div(\chi\nabla T)+\Lambda\varepsilon\\
\displaystyle
\rho\left(2\vec{\Omega}\times\vec{u}+\vec{u}\cdot\nabla\vec{u}\right)=
-\nabla p-\rho\nabla\phi_{\rm eff} +E\vec{F}_u \\
\displaystyle \div(\rho\vec{u})=0\\
p=\pi_c\lp\rho T+\frac{1-\beta_c}{\beta_c}T^4\rp\\
\end{array}
\right.
\end{equation}
where $\phi_{\rm eff}=\phi-\frac{1}{2}\Omega^2r^2\sin^2\theta$ and
$\vec{F}_u=\Delta\vec{u}+\frac{1}{3}\nabla(\div\vec{u})$. We also
introduced the dimensionless numbers

\[
E=\frac{\mu\lambda}{\rho_c R^2},\quad
\mathcal{P}=\frac{\mu\mathcal{R}_M}{\chi_c}.\]
Here, $E$ is a kind of Ekman number associated with the limit angular
velocity $\sqrt{4\pi G\rho_c}$, and $\mathcal{P}$ the Prandtl
number at the center. The Ekman number measures the importance of
viscous force relative to Coriolis force, and its true definition is
$E/\Omega$; here we had to keep out the non-dimensional rotation rate for
convenience. In stellar conditions $E/\Omega\ll1$. Other, more stellar,
dimensionless numbers have been introduced:

\[\pi_c=\frac{\mathcal{R}_MT_c\lambda^2}{R^2},\quad
\Lambda=\frac{\varepsilon_c R^2}{\chi_c T_c} \quad \mbox{and} \quad
\beta_c=\frac{\pi_c}{\pi_c+ \frac{aT_c^4\lambda^2}{3\rho_cR^2}},
\]
where $\pi_c$ is the dimensionless central gas pressure and $\beta_c$
the usual ratio of gas to total pressure at the star center.  Once we
fix the microphysics of the problem, the dimensionless numbers $\pi_c$
and $\Lambda$  should appear as part of the solution, as they determine
the central temperature and density (or mass) of the configuration.

The quantities $\chi$ and $\varepsilon$ have been scaled using their central
values $\chi_c$ and $\varepsilon_c$, so that the dimensionless radiative
diffusivity and energy generation rate are given by

\begin{equation}
\chi=T^{\beta+3}\rho^{-\eta-1}\quad\mbox{and}\quad\varepsilon=\rho^2 T^{-2/3}
\mathrm{e}^{-\alpha(T^{-1/3}-1)}
\end{equation}
where $\alpha=bT_c^{-1/3}$. Using these expressions and writing the entropy
gradient as a function of the temperature and pressure gradients, the
equation of energy balance reads

\begin{equation}
\Delta T+\nabla\ln\chi\cdot\nabla T
-\frac{\mathcal{P}}{E}\vec{a}\cdot\vec{u}=-\Lambda\frac{\rho^{\eta+3}}{T^{\beta+11/3}}
\mathrm{e}^{-\alpha\left(T^{-1/3}-1\right)}
\end{equation}
where we introduced the vector

\begin{equation}
\vec{a}=\frac{\rho^{\eta+1}}{T^{\beta+3}}\left(\frac{\gamma}{\gamma-1}\rho\nabla T-
\frac{1}{\pi_c}\nabla p\right),
\end{equation}
which makes explicit the entropy gradient; $\gamma$ is the adiabatic exponent.

The scaling of the velocity field requires some consideration. In
\cite{R06} it is shown that the meridional circulation, \ie the radial
and meridional components of the velocity field, is proportional to the
Ekman number; therefore, we write:

\begin{equation}
\vec{u}=Eu_r\vec{e}_r+Eu_\theta\vec{e_\theta}+u_\varphi\vec{e_\varphi}.
\end{equation}

The system \eq{nondim_eq} is the set of nonlinear PDE that governs the
radiative zone of a rotating star when no turbulence or magnetic field
troubles the place. It is this system that will give us the first view
of the dynamics inside the central parts of a rapidly rotating fully
radiative star.

\section{Numerical method}


One-dimensional stellar evolution codes usually use a few thousand of
grid points to model and follow the evolution of a star. With the same
kind of discretization (i.e. finite differences), the number of grid
points would grow by a factor of a few hundred for a two-dimensional
model.  This large number of points is certainly very heavy to manage;
two-dimensional models require more efficient discretization, which is
why we adopted spectral methods.

Spectral methods are best-known for their precision when the number of grid
points is limited \cite[][]{Peyret02,Grandclement06,BGM99}. On the dark side,
they have never been used in stellar modeling (to our knowledge) and are
rather ``touchy" as far as stability is concerned; more particularly,
discontinuities require special care like a multi-domain approach does.
Nevertheless, the precision they can achieve is a major advantage,
especially when the oscillation spectrum of a star needs to be computed.
The advances in observational asteroseismology, with space missions
like COROT, indeed require very precise models \cite[][]{RLR06}.

The discretization of the equations is made via an expansion of the
unknowns onto the spherical harmonic basis for the angular part and
using Gauss-Lobatto collocation nodes for the radial part. This latter
grid is associated with Chebyshev polynomials
\cite[e.g.][]{CHQZ06,fornberg}.

\subsection{Projection on the spherical harmonics}

The variables are thus first projected on the spherical harmonics. Hence,
for the scalar quantities:

\begin{equation}
f=\sum f_l(r) Y_l^0(\theta).
\end{equation}
Note that, due to the symmetry of the problem, solutions are axisymmetric
and therefore $m=0$. They are also symmetric with respect to the
equatorial plane, so scalar quantities need only spherical
harmonics of an even order. For the velocity, we follow \cite{rieu87}
and use the expansion:

\begin{equation}
\vec{u}=\sum_{l=0}^{L_{\rm max}}
Eu_l(r)\vec{R}_l^0+Ev_l(r)\vec{S}_l^0+w_l(r)\vec{T}_l^0,
\end{equation}
where the vectorial spherical harmonics are
\begin{equation}
\begin{array}{l}
\displaystyle \vec{R}_l^m=Y_l^m\vec{e}_r\\
\displaystyle \vec{S}_l^m=\frac{\partial Y_l^m}{\partial\theta}\vec{e}_\theta+
\frac{1}{\sin\theta}\frac{\partial Y_l^m}{\partial\varphi}\vec{e}_\varphi\\
\displaystyle \vec{T}_l^m=\frac{1}{\sin\theta}\frac{\partial Y_l^m}{\partial\varphi}\vec{e}_\theta
-\frac{\partial Y_l^m}{\partial\theta}\vec{e}_\varphi.
\end{array}
\end{equation}

The projection of the momentum equation yields three equations:
\begin{eqnarray}
\frac{\mathrm{d}p_l}{\mathrm{d}r}
+2\left<\rho\right>\Omega
\left[(l-1)\alpha_{l}w_{l-1}-(l+2)\alpha_{l+1}w_{l+1}\right] && \nonumber\\
-E^2\left[\frac{4}{3}\frac{\mathrm{d}^2u_l}{\mathrm{d}r^2}+
\frac{8}{3r}\frac{\mathrm{d}u_l}{\mathrm{d}r}-
\frac{(l(l+1)+8/3)}{r^2}u_l \right. &&
\nonumber\\
\left. -\frac{l(l+1)}{3r}\frac{\mathrm{d}v_l}{\mathrm{d}r}+
\frac{7l(l+1)}{3r^2}v_l\right] &=&\mathrm{rhsu}_l
\label{momu}
\end{eqnarray}

\begin{eqnarray}
\frac{p_l}{r}
+2\left<\rho\right>\Omega\left[\frac{l-1}{l}\alpha_{l}w_{l-1}+
\frac{l+2}{l+1}\alpha_{l+1}w_{l+1}\right]\nonumber \\
-E^2\left[\frac{\mathrm{d}^2v_l}{\mathrm{d}r^2}+\frac{2}{r}\frac{\mathrm{d}v_l}{\mathrm{d}r}
-\frac{4l(l+1)}{3r^2}v_l
+\frac{1}{3r}\frac{\mathrm{d}u_l}{\mathrm{d}r}+\frac{8}{3r^2}u_l\right]
=\mathrm{rhsv}_l &&
\label{momv}
\end{eqnarray}

\begin{eqnarray}
2\left<\rho\right>\Omega\left[\frac{\alpha_{l}}{l}u_{l-1}
-\frac{\alpha_{l+1}}{l+1}u_{l+1}-\frac{l-1}{l}\alpha_{l}v_{l-1}
-\frac{l+2}{l+1}\alpha_{l+1}v_{l+1}\right] \nonumber \\
-\left[\frac{\mathrm{d}^2w_l}{\mathrm{d}r^2}+\frac{2}{r}\frac{\mathrm{d}w_l}{\mathrm{d}r}
-\frac{l(l+1)}{r^2}w_l\right]
=\mathrm{rhsw}_l , &&
\label{momw}
\end{eqnarray}
where we have introduced the spherically averaged quantities

\begin{equation}
\left<f\right>=\frac{1}{4\pi}\int_{4\pi}f\mathrm{d}\Omega=
\frac{f_0(r)}{\sqrt{4\pi}}
\end{equation}
and the coupling coefficient
\begin{equation}
\alpha_l=\frac{l}{\sqrt{4l^2-1}}.
\end{equation}
The other equations yield
\begin{equation}
\frac{\mathrm{d}^2\phi_l}{\mathrm{d}r^2}+\frac{2}{r}\frac{\mathrm{d}\phi_l}{\mathrm{d}r}
-\frac{l(l+1)}{r^2}\phi_l=\rho_l
\end{equation}
\begin{equation}
\frac{\mathrm{d}u_l}{\mathrm{d}r}+\frac{2u_l}{r}-\frac{l(l+1)}{r^2}v_l=\mathrm{rhsd}_l
\label{momd}
\end{equation}
\begin{equation}
\frac{\mathrm{d}^2T_l}{\mathrm{d}r^2}+
\left(\frac{2}{r}+\frac{\mathrm{d}}{\mathrm{d}r}\left<\ln\chi\right>\right)
\frac{\mathrm{d}T_l}{\mathrm{d}r}-\frac{l(l+1)}{r^2}T_l
-\mathcal{P}\left< a_r\right>u_l
=\mathrm{rhst}_l
\label{momt}
\end{equation}
The right-hand terms are essentially non-linear terms that are
gathered in Table 1. In their expressions $\delta$ denotes the
non-spherical part of a variable, i.e.
$\delta f=f-\left<f\right>$.

\begin{table*}
\begin{center}
\fbox{\begin{minipage}[t]{17cm}

\begin{equation}
\begin{array}{rcl}
\displaystyle \vec{\mathrm{RHS}}_\mathrm{mom}&=&
\displaystyle\sum\left(\mathrm{rhsu}_l\vec{R}_l^0+\mathrm{rhsv}_l\vec{S}_l^0
+\mathrm{rhsw}_l\vec{T}_l^0\right)\\
&=&\displaystyle\left[-\rho\frac{\partial}{\partial r}\left(\phi-\frac{1}{2}\Omega^2r^2\sin^2\theta\right)
+2\delta\rho\Omega u_\varphi\sin\theta+\frac{\rho u_\varphi^2}{r}+E^2\rho\left(
\frac{u_\theta^2}{r}-u_r\frac{\partial u_r}{\partial r}-\frac{u_\theta}{r}\frac{\partial
u_r}{\partial\theta}\right)\right]\vec{e}_r+\\
&&\displaystyle\left[-\frac{\rho}{r}\frac{\partial}{\partial \theta}
\left(\phi-\frac{1}{2}\Omega^2r^2\sin^2\theta\right)
+2\delta\rho\Omega u_\varphi\cos\theta+\frac{\rho u_\varphi^2}{r}\cot\theta-E^2\rho\left(
\frac{u_\theta u_r}{r}+u_r\frac{\partial u_\theta}{\partial r}+\frac{u_\theta}{r}\frac{\partial
u_\theta}{\partial\theta}\right)\right]\vec{e}_\theta+\\
&&\displaystyle\left[-2\delta\rho\Omega\left(u_r\sin\theta+u_\theta\cos\theta\right)
-\rho\left(\frac{u_\varphi}{r}\left(u_r+u_\theta\cot\theta\right)
+u_r\frac{\partial u_\varphi}{\partial r}+\frac{u_\theta}{r}\frac{\partial u_\varphi}{\partial\theta}
\right)\right]\vec{e}_\varphi
\end{array}
\end{equation}
\begin{equation}
\mathrm{rhsd}=-\vec{u}\cdot\nabla\ln\rho
\end{equation}
\begin{equation}
\mathrm{rhst}=-\Lambda\frac{\rho^{\eta+3}}{T^{\beta+11/3}}
\mathrm{e}^{-\alpha\left(T^{-1/3}-1\right)}+\frac{\mathcal{P}}{E}\delta\vec{a}\cdot\vec{u}
-\nabla\delta\ln\chi\cdot\nabla T
\end{equation}
\end{minipage}
}
\end{center}
\caption[]{Non-linear terms that appear in the right-hand sides of the
equations of motion (\ref{momu}-\ref{momw}), \eq{momd} and \eq{momt}.
}
\end{table*}

\subsection{Boundary conditions}

Boundary conditions have to be expressed on the radial functions. 
Regularity of the solutions at the star center demands the regularity of
the radial functions at $r=0$. As far as the pressure is concerned, we
fix the polar value to $p_s$ which fixes the constant needed by the
solution of the momentum equation. Stress-free boundary conditions on
the velocity field give \cite[][]{rieu87}

\begin{equation}
\left\{\begin{array}{ll}
\displaystyle u_l=0&\\
\displaystyle r\frac{\mathrm{d}v_l}{\mathrm{d}r}-v_l=0& \\
\displaystyle r\frac{\mathrm{d}w_l}{\mathrm{d}r}-w_l=0&
\end{array}\right.
\end{equation} 
at the surface.
The gravitational potential at the surface matches that of the vacuum
\begin{equation}
r\frac{\mathrm{d}\phi_l}{\mathrm{d}r}+(l+1)\phi_l=0 \qquad\mbox{at}\quad
r=1.
\end{equation}
Finally, the radial functions of the temperature verify

\begin{equation}
r\dnr{T_l}+\sigma_T T_l=0
\end{equation}
at the surface.

\subsection{Algorithm for iterations}

Equations are solved iteratively because of nonlinearities. We follow
and generalize the algorithm used by \cite{BGM98} for polytropic stellar
models. The idea is to set the equations in the form

\[ {\cal L}_n(\vy_{n+1}) = RHS(\vy_n,\vy_{n-1})\]
where ${\cal L}_n$ is a linear operator that may depend on the n-th
iterate. Such a scheme is also known as the fixed-point algorithm.
Convergence depends on the norm of the linear operator and the closeness
to the solution; the dependence of RHS on $\vy_{n-1}$
stands for the relaxation terms. Unlike Henyey's method, which is a Newton type
algorithm, this scheme does not need the computation of the Jacobian
matrix, which is quite convenient with our discretization and high order
operators.

\section{Tests of the model}

\subsection{Internal accuracy}

\begin{figure*}
\centering
\includegraphics[width=17cm]{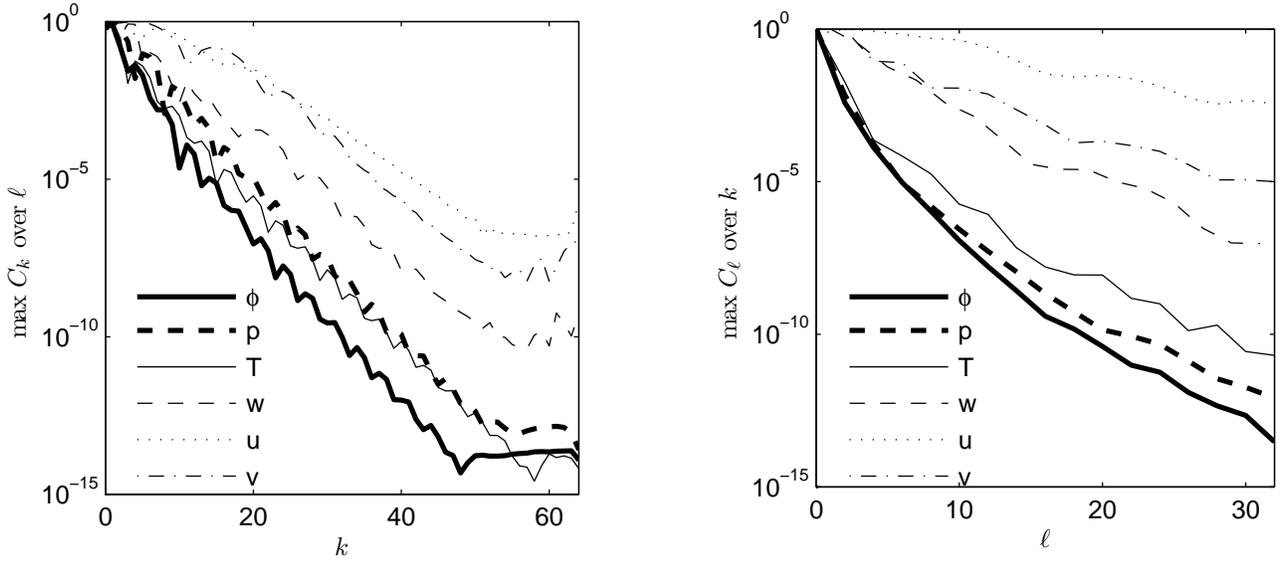}
\caption{Spectral convergence of the various physical quantities of the
model. On left: scaled coefficient of the Chebyshev polynomial expansion
(maximized over the spherical harmonic degrees); on right: the
corresponding spectral coefficients of the spherical harmonics maximized
over the Chebyshev coefficients. These curves have been obtained with a
model where $p_s=10^{-5}$, $E=10^{-8}$, $\Omega=0.07$, $\alpha=15.2$,
$\sigma_T=1.5$.}
\label{spectral}
\end{figure*}

In Fig.~\ref{spectral} we show the spectra of the different physical
quantities. As these spectra are scaled by the maximum value, they show
the internal relative precision of the solutions due to truncation. In
this example, spectral convergence is very good on the gravitational
potential, pressure, temperature and less good on the velocity
components; this last quantity is indeed facing rapid variations in
the boundary (Ekman) layers.

Although spectra give a good idea of the precision of the solution, they
do not tell the whole story. Indeed, round-off errors also affect
the solution as they are usually amplified by the resolution of linear
systems (backward error) and may have a devastating effect
\cite[e.g.][]{VRBF06}. In our case, round-off errors have been tested on
``eigenvalues" like $\pi_c$ and $\Lambda$ and were found to be below
$5\,10^{-8}$.

\subsection{The virial equality}
A standard test of the accuracy of the computations of a
rotating, self-gravitating fluid is the virial equality
\cite[see][]{OB68,EM85}. This is indeed an integral form of the
momentum equation, along with mass conservation. Since this equality
has only been used with barotropic stars, we give its full
expression here as is appropriate for a baroclinic star. 

We first rewrite the momentum and continuity equations as 

\begin{eqnarray}
(2\vO\wedge\rho\vu)_i + \rho u_j\partial_ju_i &=& 
-\rho\partial_i\phi+\rho\Omega^2s(\es)_i +
\partial_j\sigma_{ij}\label{mom}  \\
\partial_j\rho u_j &=& 0,
\end{eqnarray}
where $\sigma_{ij}$ stands for the stress tensor (including the
pressure), namely:

\[ \sigma_{ij} = -P\delta_{ij}+\mu\lp\partial_iv_j+\partial_jv_i -
\frac{2}{3}(\partial_kv_k)\delta_{ij}\rp. \]

 On the outer boundary, we imposed stress-free boundary
conditions, \ie $u_r=\sigma_{r\theta}=\sigma_{r\varphi}=0$. The volume
integral of the scalar product of \eq{mom} with $\vr$ yields the
virial identity

\[ 2T_{\rm rel}+I\Omega^2 +W + 3P + I_{s} + 2\Omega L_z = 0,\]
where we introduced

\[ T_{\rm rel} = \demi\intvol\rho u^2dV\]
as the kinetic energy in the rotating frame,

\[ I=\intvol \rho r^2\sin^2\theta dV \]
as the momentum of inertia,

\[ W = \demi\intvol\rho\phi\dV\]
as the gravitational potential energy,

\[ 3P = -\intvol \sigma_{ii}dV \]
as the internal energy,

\[I_s=\intsur r_i\sigma_{ij}dS_j \]
as the surface stress and

\[ L_z = \ez\cdot \intvol\vr\wedge\rho\vu dV \]
as the relative angular momentum z-component. The last two integrals may
be vanishingly small if the background rotation contains all the
angular momentum and if the boundary of the domain is near the
zero pressure surface.

We give in Table \ref{virial} the value of the virial equality scaled, as
usual, by the gravitational energy $W$. We have used 32 radial points,
$L_\mathrm{max}=16$, Ekman number $E=10^{-5}$ and surface pressure
$p_s=10^{-3}$. For comparison, the break-up velocity for these
models is $\Omega_k\sim0.2$. The virial equality is verified with errors
less than 10$^{-7}$ thanks to spectral precision as in \cite{BGM98}. These
values can be compared with the $10^{-3}$ obtained by \cite{UE94,UE95}
or $10^{-4}-10^{-5}$ by \cite{JMS05} resulting from the finite difference
discretizations.

\begin{table}
\centerline{
\begin{tabular}{cc}
\hline
$\Omega$&Virial\\
$10^{-8}$&$-5.4\,10^{-8}$\\
$0.05$&$-1.4\,10^{-8}$\\
$0.1$&$-1.3\,10^{-8}$\\
$0.15$&$1.4\,10^{-7}$\\
\hline
\end{tabular}
}
\caption[]{Values of the virial equality scaled by the gravitational energy 
for different rotation velocities.
}
\label{virial}
\end{table}

\subsection{Calibration with one-dimensional models}

\begin{table}
\centerline{
\begin{tabular}{ccc}
\hline
       &ESTER & CESAM \\
Mass   & 3.05 & 3.00 \\
Radius & 2.40 & 2.29 \\
Luminosity & 107. & 107. \\
$\rho_c$& 129  & 132 \\ 
$\pi_c$ & $1.39\,10^{-3}$ & $1.48\,10^{-3}$ \\
$\Lambda$ & 232  & 219\\
\hline
\end{tabular}
}
\caption[]{Comparison of the results between the two-dimensional code
ESTER run at zero rotation and the one-dimensional code CESAM.
}
\label{compar}
\end{table}

In the dimensionless formulation, the two numbers $\Lambda$ and $\pi_c$
are of special importance. They give the radius, the central
temperature, and the central density once the mass is fixed. $\pi_c$ is
indeed a non-dimensional measure of the central pressure while $\Lambda$
of central heating. Using the expression of these numbers we note that
their product

\[ \Lambda\pi_c = \frac{{\cal R}\eps_c}{4\pi G\rho_c\khi_c}\]
depends only on central density and central temperature; thus, its
determination yields a first relation between these two quantities. The
expression of the mass

\[ M= \rho_cR^3\intvol\rho dV\]
combined with that of $\Lambda$ gives another relation between $T_c$ and
$\rho_c$, thereby allowing for the determination of $R, T_c, \rho_c$.
Actually, for convenience we prefer fixing the central temperature
instead of the mass and deriving the other quantities.

\begin{figure}[t]
\includegraphics[width=8cm]{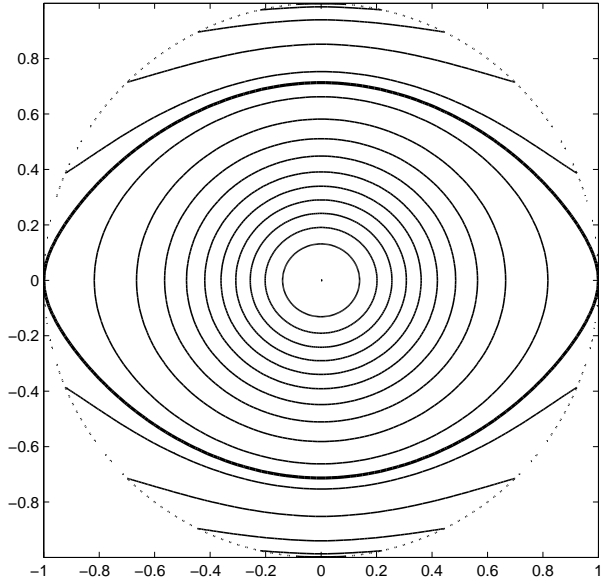}
\caption[]{Isobars for a model with $\Omega=0.07$.  An equatorial Keplerian
velocity for this case is reached when $\Omega_k=0.082$, so the star is
rotating at $85\%$ of its break-up velocity. Other parameters are
standard (see text). The thick isobar underlines the last complete
isobar allowed by the container.}
\label{isobars}
\end{figure}

\begin{figure}[t]
\includegraphics[width=8cm]{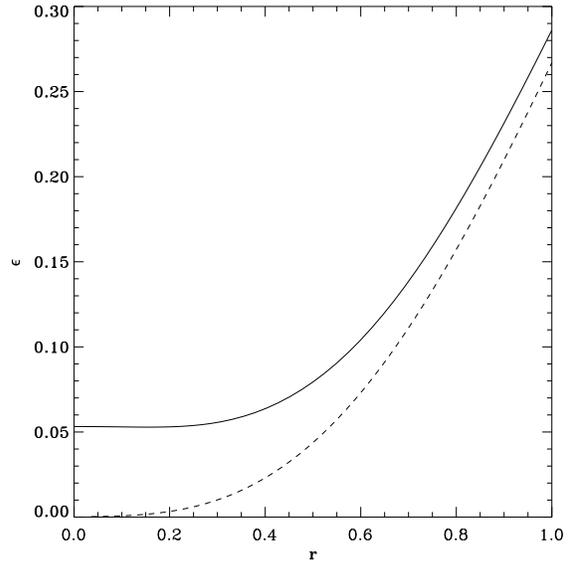}
\caption{Flatness $\epsilon=1-\frac{R_\mathrm{pol}}{R_\mathrm{eq}}$
of the isobars as a function of the equatorial radius for our model
(solid line) and for the polytrope of index 4.37 rotating at 82\% of the
critical velocity.  $p_s=10^{-5}$,
$E=10^{-8}$, $\Omega=0.07$, $\alpha=15.2$, $\sigma_T=1.5$.}
\label{epsilon}
\end{figure}

As a first test of the model, we compared the results obtained for a
non-rotating configuration with the results of a one-dimensional code like
CESAM \cite[][]{Morel97}. We show the results in Table~\ref{compar}.
There,
we used a central temperature of 3.05\,$10^7$ and a hydrogen mass
fraction of X=0.712. The physics of the one-dimensional code was made
as close to ours as possible, but surface boundary conditions, implying
some superficial convection, still make noticeable differences. (The
three-solar mass was chosen to minimize the effects of convection although
the physics does not strictly apply in this case.)  Nevertheless, the
results displayed in this table show that our model is close to models
computed with traditional stellar evolution codes.  Thus, we have a
self-gravitating fluid in our ``box", close to the stellar conditions.

We do not ask for more precision at the moment since all the
physics is not implemented; we leave for future work a more detailed
comparison. Presently, we estimate that mass and temperature distributions
are close enough to ``reality" that our fluid flows are meaningful
(as long as turbulence is negligible).

\section{Results}

For all the examples shown below, we have used a polar pressure of
$p_s=10^{-5}$ (scaled by central pressure), a rotation rate of
$\Omega=0.07$, which is about 82\% of the
critical angular velocity, a central temperature of
$T_c=(b/\alpha=15.2)^3=1.33\, 10^7$~K, $\sigma_T=1.5$ and $\beta_c=1$ (we
neglect radiation pressure). The Ekman number was set to $10^{-8}$ and
Prandtl number to zero. The mass enclosed in the container was
close to one solar mass.

Table~\ref{resum} summarizes the properties of the models when rotation
is increased. Numerics use a radial grid with Nr=64 points and the
maximum degree of spherical harmonics is L$_{\rm max}$=32. We use
$p_s=10^{-5}$, $T_c=1.3285\, 10^7$~K, $\sigma_T=1.5$ and $\beta_c=1$;
hydrogen mass fraction is X=0.71 and Z=0. Ekman number is $E=10^{-8}$.
We note that the mass increases slightly, as well as the luminosity,
while the radius and the central density decrease slightly.  This is a
consequence of our choice to keep the central temperature and the ratio
of polar pressure to central pressure constant.%

\begin{table*}
\centerline{
\begin{tabular}{cccccccc}
\hline
$\Omega$ & 0.01 & 0.02 & 0.03 & 0.04 & 0.05 & 0.06 & 0.07\\
Mass (M$_\odot$)   & 1.003 & 1.009 &1.018 & 1.032&1.050 & 1.074&1.105\\
Radius (R$_\odot$) & 0.9955& 0.9941&0.9918&0.9883&0.9833&0.9765&0.9674\\
Luminosity (L$_\odot$)&0.803&0.805&0.808 &0.811 &0.816&0.822& 0.829\\
$\rho_c$ (g/cm$^3$)& 86.5   & 86.4&86.3 & 86.2 &86.0&85.8 & 85.6 \\
$\pi_c$ & $5.16\,10^{-3}$ & $5.18\,10^{-3}$ & $ 5.21\,10^{-3}$& $5.268\,10^{-3}$ &
$5.32\,10^{-3}$ &$5.41\,10^{-3}$ &$5.53\,10^{-3}$ \\
$\Lambda$ & 69.2 & 69.4& 69.0 & 68.7 &66.9&65.5&63.7 \\
\hline
\end{tabular}
}
\caption[]{Parameters of our purely radiative star for various rotation
rates.}
\label{resum}
\end{table*}

\subsection{Internal structure}

\begin{figure}
\centering
\includegraphics[width=7cm]{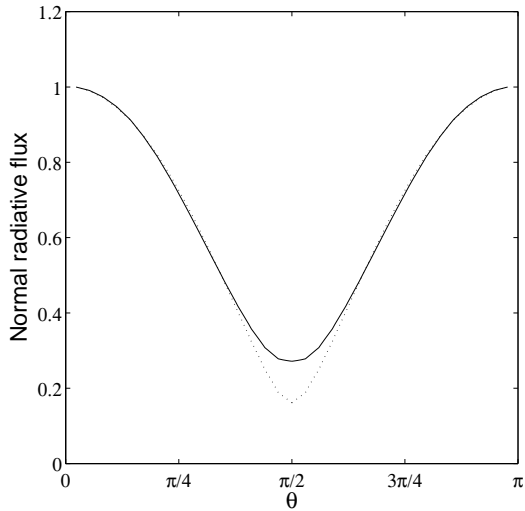}
\caption[]{Normal radiative flux at  the last isobar  ($p=p_{\rm eq}$),
real values (solid line) and the von Zeipel model (dotted line). The von Zeipel
model is calculated from the normal effective gravity; both curves have
been scaled by their polar value.}
\label{fluxprof}
\end{figure}

The first view of the internal structure is given by the pressure
distribution as shown in Fig.~\ref{isobars}. We emphasize the
``last isobar" with a thick line. This is the last complete isobar
fitting in the container, and we use it below to appreciate the
latitude variations of the flux and temperature.

Mass distribution can be appreciated by the flatness profile of the
isobars, as shown in Fig.~\ref{epsilon}. It is well known that a stellar
envelope with power law opacities behaves like a polytropic
envelope of index $n=(\beta+3)/(\eta+1)$, which is $\sim4.37$ in our
case. Thus, for comparison, we also show the flatness of the isobar of a
fully polytropic star rotating at 85\% of the breakup velocity. 
The curves show that the polytrope is much more centrally condensed
than the star, as the flatness of isobars are almost zero at the center.

We then computed the energy flux surface density at the last isobar
contained in the domain (Fig.~\ref{isobars}) and compared it to the
one given by the von Zeipel model. We recall that this model assumes
that the star is barotropic, so that density and temperature depend
only on the effective potential (\ie gravitational plus centrifugal);
it follows that the energy flux is proportional to the local effective
gravity. Figure~\ref{fluxprof} shows that the von Zeipel model overestimates
the contrast of the polar and equatorial flux. Actually, the ratio of
polar to equatorial flux is 3.68 for our model, while it is 6.2 for von
Zeipel one, almost two times higher.

\begin{figure}
\centering
\includegraphics[width=7cm]{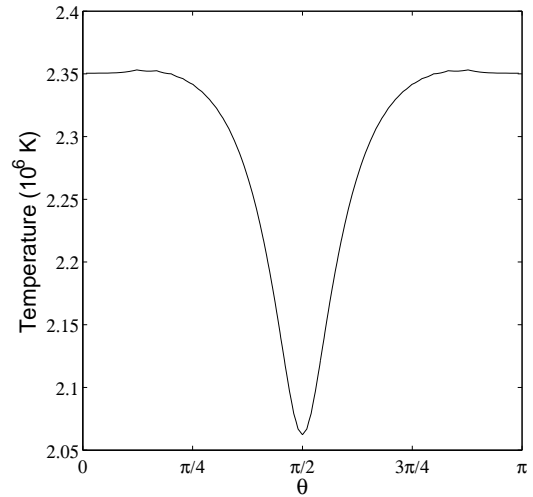}
\caption{Variations of temperature along the last isobar.}
\label{var_temp}
\end{figure}

As far as temperature is concerned, we computed the variation of
temperature along the last isobar. As shown in Fig.~\ref{var_temp}, the
temperature drops noticeably at the equator, i.e. by a few $10^5$~K.
The temperature drop shows that isothermal surfaces are more spherical
than the isobars.

\begin{figure}
\centering
\includegraphics[width=7cm]{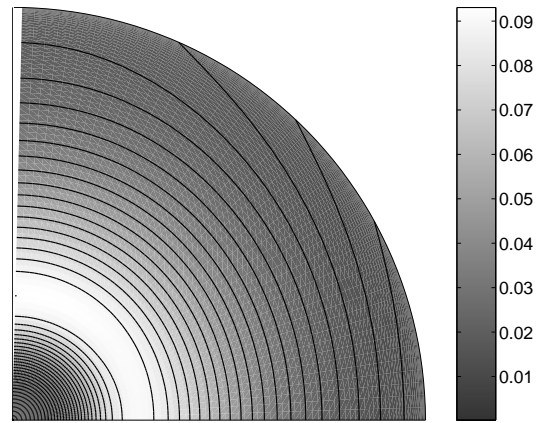}
\caption[]{Square of the \BVF\ $N^2$ when the rotation is
$\Omega=0.03$.}
\label{bv1}
\end{figure}

\begin{figure}
\centering
\includegraphics[width=7cm]{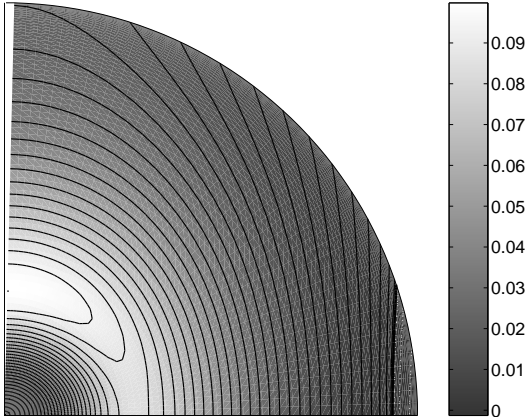}
\caption[]{Same as in Fig.~\ref{bv1} but with $\Omega=0.07$. Dotted lines
are for negative contours, in which there is convective instability. The
thick line represents the points where $N^2=0$.}
\label{bv3}
\end{figure}

\begin{figure}
\centering
\includegraphics[width=7cm]{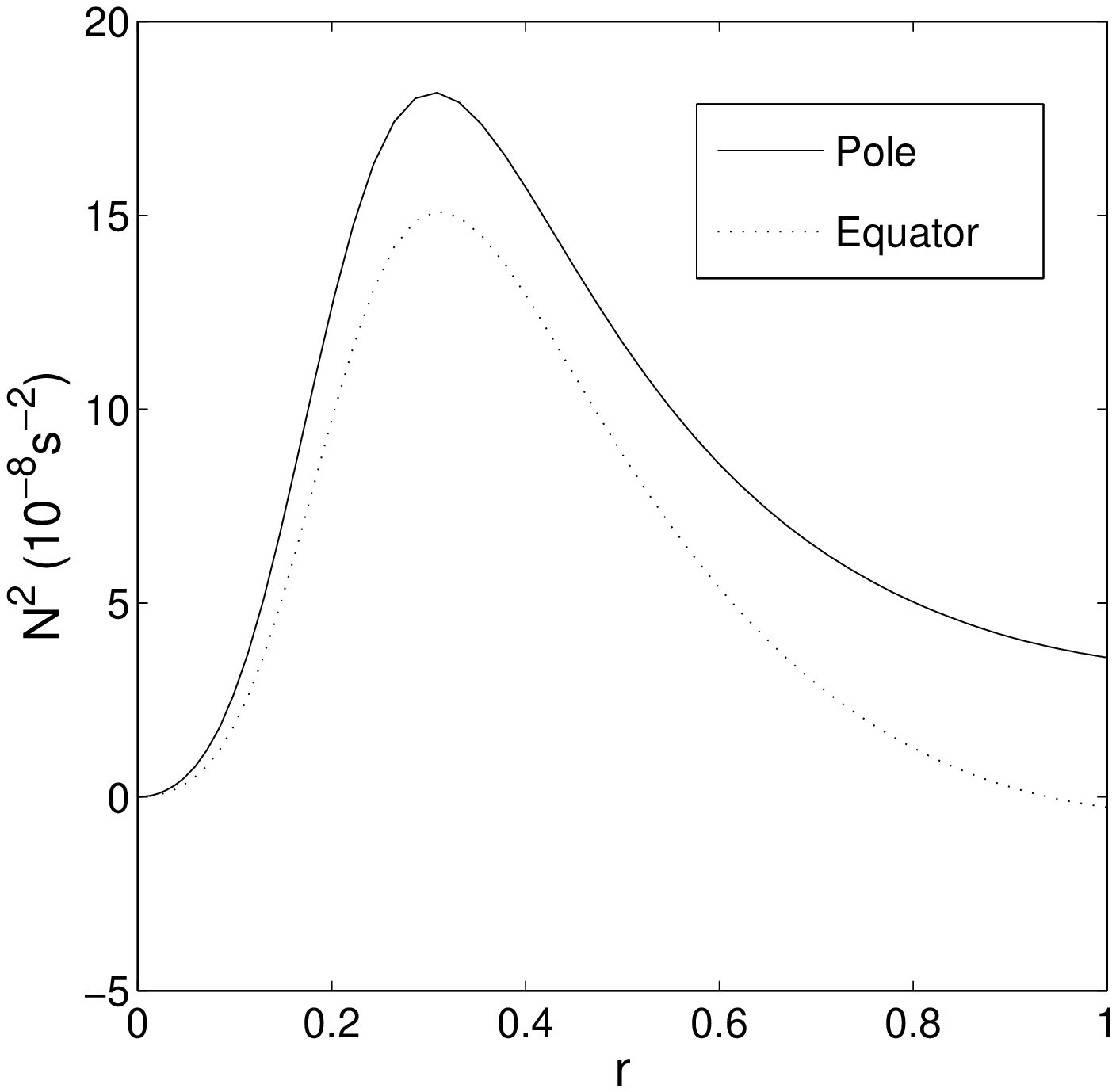}
\caption{\BVF\ profile at pole and equator when \mbox{$\Omega=0.07$}.}
\label{bvprof}
\end{figure}

Next we considered the local \BVF which, as shown in \cite{R06},
controls the baroclinic torque. This quantity is plotted in
Fig.~\ref{bv1} for a rotation of $\Omega=0.03$ and in Fig.~\ref{bv3}
for the usual rotation $\Omega=0.07$.
Quite interestingly, we see that if rotation is fast enough, the local
\BVF\ is imaginary in the equatorial region revealing the appearance of a
convection zone driven by the cool equator. Of course in such a region
the model that we have used is no longer valid as one should take into
account the entropy mixing of turbulent convection. However, we think
that we show here, for the first time, the emergence of an equatorial
convection zone due a fast rotation. In Fig.~\ref{bvprof}
we underline the anisotropy of the squared \BVF\ distribution by
comparing the poleward and equatorward profiles. This anisotropy will
impact on the low-frequency spectrum of oscillations (gravito-inertial
modes) as can be surmised from \cite{DR00}.

\subsection{Differential rotation and circulation}

\begin{figure}
\centering
\includegraphics[width=8cm]{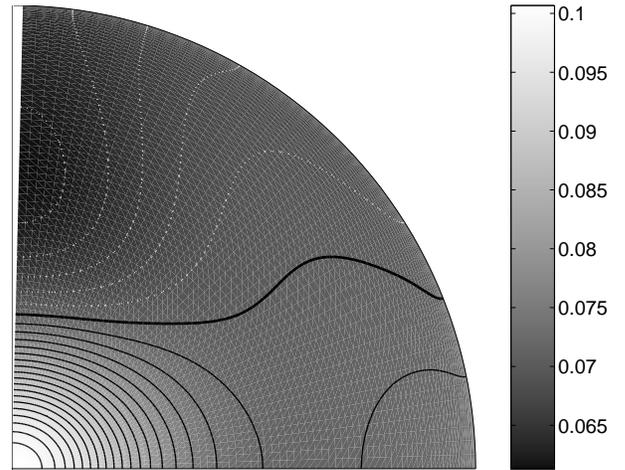}
\caption{Differential rotation. The thick line corresponds
to $\Omega=0.07$, which is the background rotation rate. Faster and slower
rotation are represented by solid black lines and dotted white lines,
respectively.}
\label{diffrot}
\end{figure}

We now turn to the dynamical state of our radiative ``star". Since we
simplified the solutions by only focusing on the axisymmetric steady
ones, the dynamics of the star is described by its differential
rotation and meridional circulation. We give in Fig.~\ref{diffrot} the
typical shape of isorotation lines in a meridian plane. These do not
change much with rotation. The main property of this baroclinic flow is
that the equator rotates faster than the pole.

This shape of the differential rotation can be understood from the
torque balance in the momentum equation. Indeed, neglecting nonlinear
and viscous terms, the curl of the momentum
equation in \eq{nondim_eq}, together with the equation of state, yields
the baroclinic balance, namely,

\[ \dz{u_\varphi} = \pi_c(\na T\times\na\ln P)_\varphi.\]
Now, if we consider the latitudinal variation of the temperature along
an isobar, replacing the radial variable by the pressure, we may write
$T\equiv T(P,\theta)$, and thus

\[ \na T = \frac{\partial T}{\partial P}\na P + \lp\frac{\partial
T}{\partial\theta}\rp_P\na\theta.\]
We can therefore rewrite the baroclinic balance as

\beq
\dz{u_\varphi} = -\frac{\pi_c}{r}\lp\frac{\partial
T}{\partial\theta}\rp_P\frac{\partial\ln P}{\partial r}.
\eeq
Since in any direction, pressure decreases with $r$ and isotherms are
more spherical than isobars (temperature decreases on an isobar from
pole to equator), we find that $\dz{u_\varphi} <0$. The baroclinic torque
thus makes the polar region rotate slower than the equatorial regions.

This result is (unfortunately!) at odds with the solution of the
Boussinesq model devised in \cite{R06}, where the polar rotation
turned out to be faster than the equatorial one.
In the Boussinesq model, the baroclinic balance leads to

\[ \dz{u_\varphi} = N^2\sth\cth,\]
which shows that rotation increases with $z$, meaning with latitude
in the northern equatorial region, when the squared \BVF\ is positive
(\ie with stable stratification).

Hence, compressibility changes the torque drastically. We traced this
difference to the fact that, in the Boussinesq model, density is only a
function of temperature; therefore, isochores are identical to isotherms,
while isochores are strongly influence by pressure in the compressible
case.

We also examined the dependence of the differential rotation on the
viscosity and find that it is independent of viscosity in the asymptotic
régime $E\ll1$. This property is also shared by the Boussinesq model, which
gives the explanation. As shown in \cite{R06}, viscosity is necessary to
determine the differential rotation since the inviscid case is degenerate:
an arbitrary function of $r\sth$ (a geostrophic flow) can always be
added to $u_\varphi$. Viscosity lifts this degeneracy by  ``choosing" the
geostrophic flow such that no tangential stress is exerted on the outer
boundary, while the meridional flow exactly matches the Ekman pumping. As
Ekman pumping and meridional flows are both proportional to viscosity,
this quantity simplifies and we are left with a differential rotation
independent of it.

\begin{figure}
\centering
\includegraphics[width=7cm]{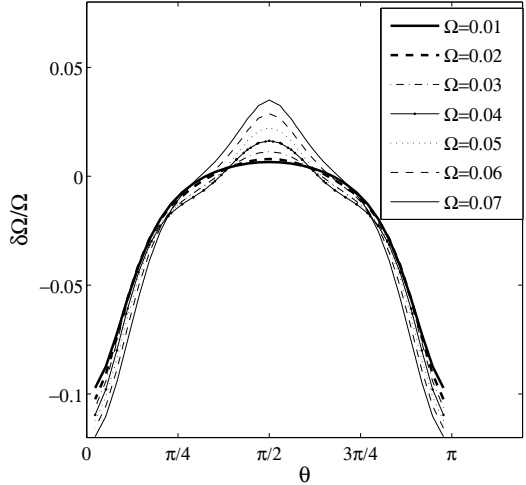}
\caption{Relative differential rotation on the last isobar.}
\label{dr_surf}
\end{figure}

Furthermore, an interesting property is shown in Fig.~\ref{dr_surf}:
except for rotations faster than $\Omega=0.03$ ( $\sim 36$\% of the
critical angular velocity), the relative surface differential rotation
keeps the same profile. Poles are 10\% slower than the equator. If the shape
of this profile is reachable through observations, a comparison to this
shape will give interesting information on the dynamical state of the
fluid (presence of turbulence, magnetic fields, etc.).

We also note from Fig.~\ref{dr_surf} that at a high rotation rate, a kind
of equatorial jet develops between latitudes $\pm20$°. If confirmed by
more realistic models, such a jet may play an important role in mass
loss for stars rotating close to the breakup limit.

\begin{figure}
\centering
\includegraphics[width=6.5cm]{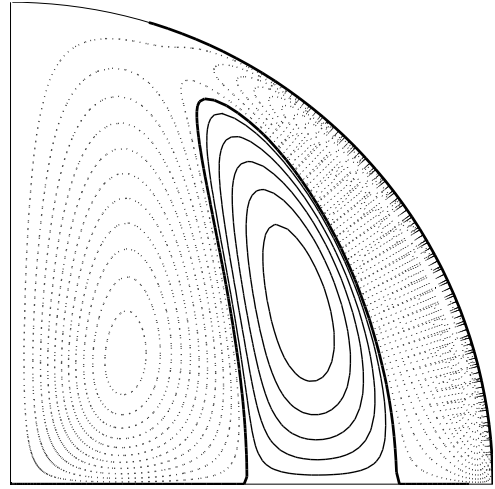}
\caption{Meridional circulation. Solid lines represent counterclockwise
circulation and dotted lines clockwise circulation.  $p_s=10^{-5}$, $E=10^{-8}$,
$\Omega=0.07$, $\alpha=15.2$, $\sigma_T=1.5$.}
\label{merid_circ}
\end{figure}

The case of meridional circulation, displayed in Fig.~\ref{merid_circ},
shows three cells of circulation. This feature of the meridional flow is
controlled by the distribution of the \BVF\ and the Prandtl number as
shown in the Boussinesq case. As far as the turnover time associated
with this circulation is concerned, we refer to the discussion in
\cite{R06}; indeed, except the profile of density, there is no
significant change, because the turnover time remains controlled by viscosity
(probably turbulent) and the Prandtl number.

\begin{figure}
\centering
\includegraphics[width=7cm]{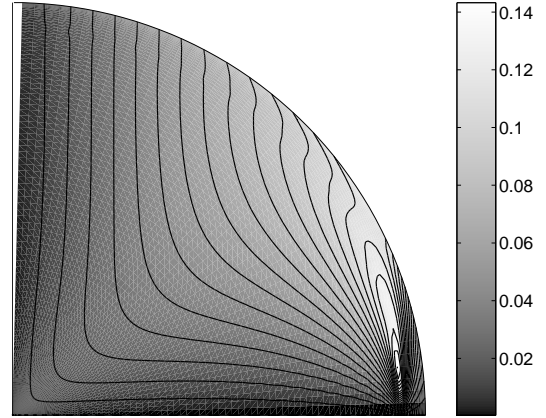}
\caption{Directional derivative of $L_z$ along equal entropy
lines, \ie $(\vn\cdot\na)L_z$ where $\vn$ is a unit vector tangential to
the isentropy surface and directed towards the equator. }
\label{dLz}
\end{figure}

Just as in the Boussinesq case questions of stability would require a full
study. As in \cite{R06}, we nevertheless take a look at the case
of axisymmetric barotropic stability. We thus compute the angular
momentum distribution along an isentropic surface; more precisely, we
verify that, along such a surface, a fluid parcel has an increasing
angular momentum when its position gets farther from the rotation axis.
This is a necessary condition for axisymmetric stability, and
Fig.~\ref{dLz} shows that this is indeed the case.

\subsection{About the effects of the container}

Finally, one may wonder what the role of the box containing our
``star'' is. Obviously, as far as the hydrostatics is concerned, the
main effect of this box is to reduce the radius (the container cuts out
the envelope).
One may also wonder about the consequences of the conflicting spherical
geometry of the box and the spheroidal shape of the mass distribution,
on the latitudinal flux variations. We find that our model gives a
milder ratio of the polar to equatorial flux than the von Zeipel model;
actually, this difference may be reduced when the container is eliminated
because its spherical geometry may soften the latitudinal variations.

We also conjecture that the effects of the container on dynamics are
fairly weak. Indeed, we note that, while changing $\Omega$ and thus the
shape of isobars and isotherms, the form of the differential rotation
remains very similar. As the flow comes essentially from the mismatch
of isobars and isotherms, which both change with respect to the sphere
as $\Omega$ varies, its invariance at low $\Omega$ pleads for a robust
result, independent of the walls of the container.

\section{Conclusions}

In this paper we presented a physically self-consistent model of a
rotating self-gravitating perfect gas heated by nuclear reactions and
enclosed in a container with a constant absorption coefficient radiating
like a black body; heat transport in the gas is purely radiative,
insured by a power law opacity.

This is a simplified model for a completely radiative star that rotates
at a significant fraction of the breakup angular velocity. We thus
generalized the approach of \cite{R06} by taking more
realistic properties of a stellar plasma into account.
Such a fluid is never in hydrostatic equilibrium and a baroclinic flow
develops. Taking viscosity into account, we computed such a flow in
its steady state, namely the differential rotation and the meridional
circulation. Interestingly enough, the results show that equatorial
regions are rotating faster than polar ones and, provided that rotation
is less than $\sim$36\% of the breakup velocity, the surface profile
$\delta\Omega/\Omega$ is independent of both viscosity and rotation rate.

Although our model is still preliminary, we had a look at the anisotropy
of the emitted flux to find that the pole/equator ratio is
less contrasted than what is predicted by the von Zeipel model. We also
showed that, when rotation is fast enough, equatorial regions become
convectively unstable.

In parallel to constructing this model, we tested - for the first
time to our knowledge - the use of spectral methods in the stellar
structure computations. Our first results are promising in terms of
precision and robustness, since we recover the precision of a finite
difference type model with ten times less grid points.
Such a gain in efficiency is necessary for all future computations of the
evolution of a rotating star with a two-dimensional model.

The next step is of course to confirm these results using a model with
coordinates adapted to the spheroidal shape of the isobars. In this case
boundary conditions can be cleanly applied, and the surface pressure can
be decreased to very low values adapted to matching an atmosphere
model. Of course, including modeled convection zones is also a
necessary step for the completeness of models to be used in the
interpretation of observations.

\begin{acknowledgements}
We thank  Bernard Pichon for providing us with 1D models from the CESAM
code and for many helpful discussions.  This work is part of the ESTER
project aimed at modeling stars in two dimensions. It is supported
by the Programme National de Physique Stellaire along with the Action
Spécifique pour la Simulation Numérique en Astrophysique.  FE thanks
the CNRS postdoctoral program for its support during his stay at the LATT
in Toulouse.

\end{acknowledgements}

\bibliographystyle{aa}
\bibliography{../../biblio/bibnew}

\begin{thebibliography}{28}
\expandafter\ifx\csname natexlab\endcsname\relax\def\natexlab#1{#1}\fi

\bibitem[{{Aufdenberg} {et~al.}(2006){Aufdenberg}, {M{\'e}rand}, {Coudé du
  Foresto}, {Absil}, {Di Folco}, {Kervella}, {Ridgway}, {Berger}, {Brummelaar},
  {McAlister}, {Sturmann}, {Sturmann}, \& {Turner}}]{Aufdenbergetal06}
{Aufdenberg}, J.~P., {M{\'e}rand}, A., {Coudé du Foresto}, V., {et~al.} 2006,
  \apj, 645, 664

\bibitem[{Bonazzola {et~al.}(1998)Bonazzola, Gourgoulhon, \& Marck}]{BGM98}
Bonazzola, S., Gourgoulhon, E., \& Marck, J.-A. 1998, Phys. Rev. D, 58, 104020

\bibitem[{Bonazzola {et~al.}(1999)Bonazzola, Gourgoulhon, \& Marck}]{BGM99}
Bonazzola, S., Gourgoulhon, E., \& Marck, J.-A. 1999, J. Computational and
  Applied Math., 109, 433

\bibitem[{Canuto {et~al.}(2006)Canuto, Hussaini, Quarteroni, \& Zang}]{CHQZ06}
Canuto, C., Hussaini, M.~Y., Quarteroni, A., \& Zang, T.~A. 2006, {Spectral
  Methods: Fundamentals in Single Domains} (Springer Verlag)

\bibitem[{{Christensen-Dalsgaard} \& {Reiter}(1995)}]{CDGR95}
{Christensen-Dalsgaard}, J. \& {Reiter}, J. 1995, in ASP Conf. Ser. 76: GONG
  1994. Helio- and Astro-Seismology from the Earth and Space, ed. R.~K.
  {Ulrich}, E.~J. {Rhodes}, Jr., \& W.~{Dappen}, 136--+

\bibitem[{Dintrans \& Rieutord(2000)}]{DR00}
Dintrans, B. \& Rieutord, M. 2000, A \& A, 354, 86

\bibitem[{{Domiciano de Souza} {et~al.}(2005){Domiciano de Souza}, Kervella,
  Jankov, Vakili, Ohishi, Nordgren, \& Abe}]{DKJVONA05}
{Domiciano de Souza}, A., Kervella, P., Jankov, S., {et~al.} 2005, A \& A, 442,
  567

\bibitem[{{Domiciano de Souza} {et~al.}(2002){Domiciano de Souza}, {Vakili},
  {Jankov}, {Janot-Pacheco}, \& {Abe}}]{DVJJA02}
{Domiciano de Souza}, A., {Vakili}, F., {Jankov}, S., {Janot-Pacheco}, E., \&
  {Abe}, L. 2002, \aap, 393, 345

\bibitem[{{Eriguchi} \& {M\"uller}(1985)}]{EM85}
{Eriguchi}, Y. \& {M\"uller}, E. 1985, \aap, 146, 260

\bibitem[{Fornberg(1998)}]{fornberg}
Fornberg, B. 1998, A practical guide to pseudospectral methods (Cambridge
  University Press)

\bibitem[{Grandclément(2006)}]{Grandclement06}
Grandclément, P. 2006, in {Stellar fluid dynamics and numerical simulations:
  from the Sun to neutron stars}, ed. M.~Rieutord \& B.~Dubrulle (EDP
  Sciences), 153--180

\bibitem[{{Jackson} {et~al.}(2005){Jackson}, {MacGregor}, \&
  {Skumanich}}]{JMS05}
{Jackson}, S., {MacGregor}, K.~B., \& {Skumanich}, A. 2005, Astrophys. J. Supp.
  Ser., 156, 245

\bibitem[{{Kervella} \& {Domiciano de Souza}(2006)}]{KDS06}
{Kervella}, P. \& {Domiciano de Souza}, A. 2006, \aap, 453, 1059

\bibitem[{{McAlister} {et~al.}(2005){McAlister}, {ten Brummelaar}, {Gies},
  {Huang}, {Bagnuolo}, {Shure}, {Sturmann}, {Sturmann}, {Turner}, {Taylor},
  {Berger}, {Baines}, {Grundstrom}, {Ogden}, {Ridgway}, \& {van
  Belle}}]{McAlisteretal05}
{McAlister}, H.~A., {ten Brummelaar}, T.~A., {Gies}, D.~R., {et~al.} 2005,
  \apj, 628, 439

\bibitem[{Morel(1997)}]{Morel97}
Morel, P. 1997, A \& A Suppl. Ser., 124, 597

\bibitem[{{Ostriker} \& {Bodenheimer}(1968)}]{OB68}
{Ostriker}, J.~P. \& {Bodenheimer}, P. 1968, \apj, 151, 1089

\bibitem[{Peterson {et~al.}(2006a)Peterson, Hummel, Pauls, Armstrong, Benson,
  Gilbreath, Hindsley, Hutter, Johnston, Mozurkewich, \&
  Schmitt}]{petersonetal06a}
Peterson, D., Hummel, C., Pauls, T., {et~al.} 2006a, ApJ, 636, 1087

\bibitem[{{Peterson} {et~al.}(2006b){Peterson}, {Hummel}, {Pauls}, {Armstrong},
  {Benson}, {Gilbreath}, {Hindsley}, {Hutter}, {Johnston}, {Mozurkewich}, \&
  {Schmitt}}]{petersonetal06b}
{Peterson}, D.~M., {Hummel}, C.~A., {Pauls}, T.~A., {et~al.} 2006b, Nature,
  440, 896

\bibitem[{Peyret(2002)}]{Peyret02}
Peyret, R. 2002, Spectral methods for incompressible viscous flow (New-York:
  Springer)

\bibitem[{{Reese} {et~al.}(2006){Reese}, {Lignières}, \& {Rieutord}}]{RLR06}
{Reese}, D., {Lignières}, F., \& {Rieutord}, M. 2006, A \& A, 455, 621

\bibitem[{Rieutord(1987)}]{rieu87}
Rieutord, M. 1987, Geophys. Astrophys. Fluid Dyn., 39, 163

\bibitem[{Rieutord(2006)}]{R06}
Rieutord, M. 2006, A \& A, 451, 1025

\bibitem[{{Roxburgh}(2004)}]{Roxburgh04}
{Roxburgh}, I.~W. 2004, A \& A, 428, 171

\bibitem[{{Uryu} \& {Eriguchi}(1994)}]{UE94}
{Uryu}, K. \& {Eriguchi}, Y. 1994, MNRAS, 269, 24

\bibitem[{{Uryu} \& {Eriguchi}(1995)}]{UE95}
{Uryu}, K. \& {Eriguchi}, Y. 1995, MNRAS, 277, 1411

\bibitem[{Valdettaro {et~al.}(2007)Valdettaro, Rieutord, Braconnier, \&
  Fraysse}]{VRBF06}
Valdettaro, L., Rieutord, M., Braconnier, T., \& Fraysse, V. 2007, J. Comput.
  and Applied Math., 205, 382, physics/0604219

\bibitem[{Zahn(1992)}]{zahn92}
Zahn, J.-P. 1992, A \& A, 265, 115

\bibitem[{{Zorec} {et~al.}(2005){Zorec}, {Domiciano de Souza}, {Fr{\'e}mat}, \&
  {Vakili}}]{Zorecetal05}
{Zorec}, J., {Domiciano de Souza}, A., {Fr{\'e}mat}, Y., \& {Vakili}, F. 2005,
  in SF2A-2005: Semaine de l'Astrophysique Francaise, ed. F.~{Casoli},
  T.~{Contini}, J.~M. {Hameury}, \& L.~{Pagani}, 363--+

\end{thebibliography}

\end{document}